\documentclass[twocolumn,superscriptaddress,showpacs,preprintnumbers,amsmath,amssymb]{revtex4}

\usepackage{graphicx}
\usepackage{dcolumn}
\usepackage{bm}
\usepackage{color} 

\begin{document}

\title{Density-dependent deformed relativistic Hartree-Bogoliubov theory in continuum}

\author{Ying Chen}
 \affiliation{State Key Laboratory of Nuclear Physics and Technology, School of Physics, Peking University, Beijing 100871, China}

\author{Lulu Li}
 \affiliation{Beijing Institute of Applied Physics and Computational Mathematics, Beijing 100088, China}
 \affiliation{State Key Laboratory of Nuclear Physics and Technology, School of Physics, Peking University, Beijing 100871, China}

\author{Haozhao Liang}
 \affiliation{State Key Laboratory of Nuclear Physics and Technology, School of Physics, Peking University, Beijing 100871, China}

\author{Jie Meng\footnote{Email: mengj@pku.edu.cn}}
 \affiliation{State Key Laboratory of Nuclear Physics and Technology, School of Physics, Peking University, Beijing 100871, China}
 \affiliation{School of Physics and Nuclear Energy Engineering, Beihang University,
              Beijing 100191, China}
 \affiliation{Department of Physics, University of Stellenbosch, Stellenbosch 7602, South Africa}

\date{\today}

\begin{abstract}
The deformed relativistic Hartree-Bogoliubov theory in continuum with the density-dependent meson-nucleon couplings is developed.
The formulism is briefly presented with the emphasis on handling the density-dependent couplings, meson fields, and potentials in axially deformed system with partial wave method.
Taking the neutron-rich nucleus $^{38}$Mg as an example, the newly developed code is verified by the spherical relativistic continuum Hartree-Bogoliubov calculations, where only the spherical components of the densities are considered.
When the deformation is included self-consistently, it is shown that the spherical components of density-dependent coupling strengths are dominant, while the contributions from low-order deformed components are not negligible.
\end{abstract}


\maketitle
\date{today}


Since the experimental discovery of a large neutron radius in $^{11}$Li~\cite{Tanihata1985PRL}, the exotic nuclear halo phenomenon becomes one of the most interesting topics close to the nucleon drip lines. In order to describe the halo phenomenon, the asymptotic behavior of nuclear densities at large distance from the center must be treated properly, and the discrete bound states, the continuum, and the coupling between them need to be dealt with simultaneously in a self-consistent way.

During the past decades, the covariant density functional theory (CDFT) has achieved great success in describing lots of nuclear phenomena in both stable and exotic nuclei~\cite{Serot1986ANP,Ring1996PPNP,Vretenar2005Phys.Rep.101,Meng2006Prog.Part.Nucl.Phys.470}, including the recent achievements in nuclear magnetic moments~\cite{Li2011Sci.ChinaPhys.Mech.Astron.204}, pseudospin symmetry~\cite{Ginocchio2005Phys.Rep.165,Liang2011Phys.Rev.C041301}, low-lying excitations~\cite{Niksic2011PPNP,Li2009Phys.Rev.C054301,Yao2011Phys.Rev.C014308}, magnetic and antimagnetic rotation~\cite{Zhao2011Phys.Lett.B181,Yu2010PhysRevC.85.024318,Zhao2012PhysRevLett.107.122501}, collective vibration~\cite{Paar2007Rep.Prog.Phys.691,Liang2008Phys.Rev.Lett.122502,Paar2009Phys.Rev.Lett.032502,Niu2009Phys.Lett.B315,Liang2009Phys.Rev.C064316}, and so on.

In particular, great efforts have been dedicated to developing the relativistic Hartree-Bogoliubov (RHB) ~\cite{Meng1996Phys.Rev.Lett.3963,Poschl1997PRL,Meng1998Phys.Rev.Lett.460,Meng1998Nucl.Phys.A3,Meng2002Phys.Rev.C041302} and relativistic Hartree-Fock-Bogoliubov (RHFB)~\cite{Long2010Phys.Rev.C031302} theories in continuum for a self-consistent description of spherical halo nuclei.
In order to describe the halo phenomena in deformed nuclei, a deformed RHB theory in continuum has been developed recently~\cite{Zhou2006AIPConf.Proc.90,Zhou2008ISPUN2007,Zhou2010Phys.Rev.C011301R,Li2012PhysRevC.85.024312,Li2012CPL}. An interesting shape decoupling between the core and halo in ${^{42,44}}$Mg
has been found~\cite{Zhou2010Phys.Rev.C011301R,Li2012PhysRevC.85.024312}.
In these applications, the deformed RHB equations are solved in a Woods-Saxon basis~\cite{Zhou2003PRC} with the partial wave method, and only nonlinear meson self-coupling interactions are used so far.

In recent years, the RHB models with the density-dependent meson-nucleon couplings have attracted more and more attention owing to improved descriptions of the equation of state at high density, the asymmetric nuclear matter, and the isovector properties of nuclei far from stability~\cite{
Brockmann1992Phys.Rev.Lett.3408,Fuchs1995Phys.Rev.C3043,Typel1999Nucl.Phys.A331,Niksic2002Phys.Rev.C024306,Long2004Phys.Rev.C034319,Lalazissis2005Phys.Rev.C024312}. Therefore, it's necessary to develop a density-dependent deformed relativistic Hartree-Bogoliubov (DDDRHB) theory in continuum.

In this brief report, the deformed RHB theory in continuum with density-dependent meson-nucleon couplings is developed for a wider compatibility of modern functionals. The key technique here is to handle the density-dependent couplings and the meson fields as well as the potentials in a deformed system with the partial wave method.

The starting point of the CDFT is a Lagrangian density where nucleons as Dirac spinors interact with each other by exchanging effective mesons ($\sigma$, $\omega$, and $\rho$) and photons~\cite{Serot1986ANP,Ring1996PPNP,Vretenar2005Phys.Rep.101,Meng2006Prog.Part.Nucl.Phys.470},
  \begin{eqnarray}
   \mathcal{L}&=&\bar{\psi}(i\gamma_{\mu}\partial^{\mu}-M)\psi+\frac{1}{2}\partial_{\mu}\sigma \partial^{\mu}\sigma-
                 \frac{1}{2}m_{\sigma}^{2}\sigma^{2}-g_{\sigma}\bar{\psi}\sigma\psi \nonumber\\
              & &-\frac{1}{4}\Omega_{\mu\nu}\Omega^{\mu\nu}+\frac{1}{2}m_{\omega}^{2}\omega_{\mu}\omega^{\mu}
                 -g_{\omega}\bar{\psi}\gamma_{\mu}\omega^{\mu}\psi\nonumber\\
              & &-\frac{1}{4}\vec{R}_{\mu\nu}\cdot\vec{R}^{\mu\nu}+\frac{1}{2}m_{\rho}^{2}\vec{\rho}_{\mu}\cdot\vec{\rho}^{\mu}
                 -g_{\rho}\bar{\psi}\gamma_{\mu}\vec{\rho}^{\mu}\cdot \vec{\tau}\psi\nonumber\\
              & &-\frac{1}{4}F_{\mu\nu}F^{\mu\nu}-e\bar{\psi}\gamma_{\mu}A^{\mu}\frac{1-\tau_{3}}{2}\psi, \label{eq1}
  \end{eqnarray}
where $M$ is the nucleon mass, and $m_{\sigma}, g_{\sigma}, m_{\omega}, g_{\omega}, m_{\rho}, g_{\rho}$ are the meson masses and density-dependent coupling strengths of the respective mesons. The field tensors for the vector mesons and photons are
\begin{eqnarray}
\Omega^{\mu\nu}&=&\partial^{\mu}\omega^{\nu}-\partial^{\nu}\omega^{\mu},\nonumber\\
\vec{R}^{\mu\nu}&=&\partial^{\mu}\vec{\rho}^{\nu}-\partial^{\nu}\vec{\rho}^{\mu},\nonumber\\
F^{\mu\nu}&=&\partial^{\mu}A^{\nu}-\partial^{\nu}A^{\mu}.
\end{eqnarray}
Following the formalism in Ref.~\cite{Ring1991ZPA}, one can derive the RHB equation,
  \begin{eqnarray}\label{eq:RHB}
    \left(\begin{array}{cc}
       h_D-\lambda & \Delta\\
       -\Delta^*   & -h^*_D+\lambda
    \end{array}\right)
    \left(\begin{array}{c}
       U_k\\
       V_k
    \end{array}\right)
    =E_k\left(\begin{array}{c}
       U_k\\
       V_k
    \end{array}\right),
  \end{eqnarray}
with the quasiparticle energy $E_k$, Fermi surface $\lambda$, and Dirac Hamiltonian
  \begin{equation}
     h_D=\mathbf{\alpha}\cdot\mathbf{p}+V(\mathbf{r})+\beta[M+S(\mathbf{r})]+\Sigma_{\rm{rear}}(\mathbf{r}),
  \end{equation}
where the Dirac matrices $\mathbf{\alpha}=\gamma^0\mathbf{\gamma}$ and $\beta = \gamma^0$. The scalar and vector potentials are respectively,
  \begin{subequations}\label{eq:potential}
  \begin{eqnarray}
     S(\mathbf{r})&=&g_{\sigma}\sigma(\mathbf{r}), \\
     V(\mathbf{r})&=&g_{\omega}\omega_{0}(\mathbf{r})+g_{\rho}\tau_3\rho_{0}(\mathbf{r})+e\frac{(1-\tau_{3})}{2}A_{0}(\mathbf{r}),
  \end{eqnarray}
  \end{subequations}
and the rearrangement term is
  \begin{equation}\label{eq:rearrangement}
     \Sigma_{\rm{rear}}(\mathbf{r})=\frac{\partial g_{\sigma}}{\partial \rho_{v}}\rho_s(\mathbf{r})\sigma(\mathbf{r})+\frac{\partial g_{\omega}}{\partial \rho_{v}}\rho_v(\mathbf{r})\omega_{0}(\mathbf{r})
                                +\frac{\partial g_{\rho}}{\partial \rho_{v}}\rho_{3}(\mathbf{r})\rho_{0}(\mathbf{r}),
  \end{equation}
which comes from the density-dependent behaviors of the meson-nucleon couplings~\cite{Niksic2002Phys.Rev.C024306,Long2004Phys.Rev.C034319},
  \begin{eqnarray}\label{eq:strength}
     g_{\phi}(\rho_v)=\left\{
     \begin{array}{ll}
     g_{\phi}(\rho_{\rm{sat}})a_{\phi}\frac{1+b_{\phi}(x+d_{\phi})^2}{1+c_{\phi}(x+d_{\phi})^2} & \rm{for}~ {\phi} = \sigma, \omega, \\
     g_{\phi}(\rho_{\rm{sat}})\exp[-a_{\phi}(x-1)] & \rm{for}~ {\phi} = \rho,
     \end{array}\right.
  \end{eqnarray}
where $\rho_{\rm{sat}}$ denotes the baryonic saturation density of nuclear matter and $x=\rho_{v}/\rho_{\rm{sat}}$.

The equations of motion for mesons and photons read
  \begin{equation}
     (-\Delta + m_{\phi})\phi(\mathbf{r})=s_{\phi}(\mathbf{r}), \label{eq:meson}
  \end{equation}
where $m_{\phi}$ are the meson masses for $\phi=\sigma,~\omega,~\rho$ and zero for the photons. The corresponding source terms are
  \begin{eqnarray}\label{eq:sourceterm}
     s_{\phi}(\mathbf{r})=\left\{\begin{array}{ll}
     -g_{\sigma}(\rho_{v})\rho_s(\mathbf{r}),  & \rm{for ~ the}~ \sigma ~ \rm{ field},\\
     ~~g_{\omega}(\rho_{v})\rho_v(\mathbf{r}), & \rm{for~ the} ~ \omega~\rm{ field},\\
     ~~g_{\rho}(\rho_{v})\rho_3(\mathbf{r}),   & \rm{for~ the} ~ \rho~ \rm{field},\\
     ~~e\rho_c(\mathbf{r}),                    & \rm{for~ the~ Coulomb~ field}
     \end{array}\right.
  \end{eqnarray}
with various densities
  \begin{subequations}\label{eq:density}
  \begin{eqnarray}
    \rho_s(\mathbf{r})&=&\sum_{k>0}V^{\dag}_{k}(\mathbf{r})\gamma_0V_k(\mathbf{r}),\label{eq11}\\
    \rho_v(\mathbf{r})&=&\sum_{k>0}V^{\dag}_{k}(\mathbf{r})V_k(\mathbf{r}),\label{eq5}\\
    \rho_3(\mathbf{r})&=&\sum_{k>0}V^{\dag}_{k}(\mathbf{r})\tau_3V_k(\mathbf{r}),\label{eq12}\\
    \rho_c(\mathbf{r})&=&\sum_{k>0}V^{\dag}_{k}(\mathbf{r})\frac{1-\tau_3}{2}V_k(\mathbf{r}),\label{eq13}
  \end{eqnarray}
  \end{subequations}
calculated within the no-sea approximation. The sum over $k>0$ runs over the quasiparticle states corresponding to single-particle energies in and above the Fermi sea.

For axially deformed nuclei with spatial reflection symmetry, the potentials in Eqs.~(\ref{eq:potential}) and (\ref{eq:rearrangement}), coupling strengths in Eq.~(\ref{eq:strength}), meson fields in Eq.~(\ref{eq:meson}), and densities in Eqs.~(\ref{eq:density}) are expanded in terms of the Legendre polynomials~\cite{Price1987PRC},
  \begin{eqnarray}
     f(\mathbf{r})=\sum_{\lambda}f_{\lambda}(r)P_{\lambda}(\cos\theta), ~ \lambda=0,2,4,\cdots
  \end{eqnarray}
with
  \begin{eqnarray}
     f_{\lambda}(r)=\frac{2\lambda+1}{2}\int_{-1}^1 d(\cos\theta) f(\mathbf{r})P_{\lambda}(\cos\theta).
  \end{eqnarray}

According to the partial wave method, first, the densities in Eqs.~(\ref{eq:density}) are represented as
  \begin{eqnarray}
     \rho(\mathbf{r})=\sum_{\lambda}\rho_{\lambda}(r)P_{\lambda}(\cos\theta).
  \end{eqnarray}
Second, the coupling strengths in Eq.~(\ref{eq:strength}) read
  \begin{equation}\label{eq:gsig}
     g_{\phi}(\rho_v)= \sum_{\lambda}g_{\phi,\lambda}(r)P_{\lambda}(\cos\theta).
  \end{equation}
By taking the $\sigma$ meson as an example,
  \begin{align}
     g_{\sigma,\lambda}(r)=&\frac{2\lambda+1}{2}\int_{-1}^{1}d(\cos\theta)P_{\lambda}(\cos\theta)g_{\sigma}(\rho_{\rm{sat}})\nonumber\\
                            &\times a_{\sigma}\frac{1+b_{\sigma}\{[\sum_{\lambda1 } \rho_{v,\lambda1}(r)
                             P_{\lambda1}(\cos\theta)]/\rho_{\rm{sat}}+d_{\sigma}\}^2}{1+c_{\sigma}\{[\sum_{\lambda2} \rho_{v,\lambda2}(r)P_{\lambda2}(\cos \theta)]/\rho_{\rm{sat}}+d_{\sigma}\}^2},\nonumber\\
  \end{align}
as illustrated for the neutron-rich nucleus $^{38}$Mg in Fig.~\ref{fig3} below. Third, the meson and Coulomb fields are, respectively, solved by the Klein-Gordon and Poisson equations~(\ref{eq:meson}), i.e.,
  \begin{eqnarray}
    \phi(\mathbf{r})&=& \int d\mathbf{r}' D(r,\theta,r',\theta';m_{\phi})s_{\phi}(r',\theta'),\nonumber\\
                    &=& \sum_{\lambda}\phi_{\lambda}(r)P_{\lambda}(\cos\theta),
  \end{eqnarray}
with
  \begin{eqnarray}
     \phi_{\lambda}(r)&=& -4\pi m_{\phi}\left[ h_{\lambda}(im_{\phi}r)\int_{0}^r dr' r'^2 j_{\lambda}(im_{\phi}r')
                            s_{\phi,\lambda}(r')\right.\nonumber\\
                      & &+\left. j_{\lambda}(im_{\phi}r)\int_{r}^{\infty} dr' r'^2 h_{\lambda}(im_{\phi}r')s_{\phi,\lambda}(r')\right]
  \end{eqnarray}
for the meson fields, where $j_{\lambda}$ and $h_{\lambda}$ are the spherical Bessel and Hankel functions in static Green functions $D(r,\theta,r',\theta';m_{\phi})$, and
  \begin{eqnarray}
     \phi_{\lambda}(r)&=&\frac{1}{r^{\lambda+1}}\int_{0}^rdr'r'^{\lambda}s_{\phi,\lambda}(r')+r^{\lambda}\int_{r}^{\infty}dr'\frac{1}{r'^{\lambda+1}}s_{\phi,\lambda}(r')\nonumber\\
  \end{eqnarray}
for the Coulomb field.
Finally, the scalar potential $S(\mathbf{r})$ in Eqs.~(\ref{eq:potential}) is written as
  \begin{eqnarray}
     S(\mathbf{r})=\sum_{\lambda}S_{\lambda}(r)P_{\lambda}(\cos\theta)
  \end{eqnarray}
with
  \begin{eqnarray}
     S_{\lambda}(r)&=&\frac{2\lambda+1}{2}\sum_{\lambda1\lambda2}g_{\sigma,\lambda1}(r)\sigma_{\lambda2}(r) \nonumber\\
                   & &\int_{-1}^1 d(\cos\theta)P_{\lambda1}(\cos\theta)P_{\lambda2}(\cos\theta)P_{\lambda}(\cos\theta),\nonumber\\
  \end{eqnarray}
where the partial waves $\lambda_1$ and $\lambda_2$ are coupled to $\lambda$. The vector potential $V(\mathbf{r})$ is obtained in the similar way.
It should also be noted that the rearrangement term in Eq.~(\ref{eq:rearrangement}),
  \begin{eqnarray}
     \Sigma_{\rm{rear}}(\mathbf{r})&=&\sum_{\lambda}\Sigma_{\rm{rear},\lambda}(r)P_{\lambda}(\cos\theta),
  \end{eqnarray}
is a result of the couplings among three partial waves.

The total energy of a nucleus is composed of
  \begin{eqnarray}
     E_{\rm{tot}}&=&E_{\rm{part}}+E_{\sigma}+E_{\omega}+E_{\rho}+E_{\rm{rear}}+E_{\rm{coul}}+E_{\rm{cm}},\nonumber\\
  \end{eqnarray}
where the center-of-mass correction $E_{\rm{cm}}$ can be calculated in an empirical or microscopic way~\cite{Bender2000Eur.Phys.J.A467,Long2004Phys.Rev.C034319,Zhao2009Chin.Phys.Lett.112102,Li2012PhysRevC.85.024312, Zhao2010Phys.Rev.C054319}.

In present Brief Report, we use the functional PKDD developed in Ref.~\cite{Long2004Phys.Rev.C034319} for the particle-hole channel and a zero range density-dependent pairing force,
  \begin{eqnarray}\label{eq:pairing}
     V^{\emph{pp}}(\mathbf{r}, \mathbf{r}')=\frac{V_0}{2}(1-P^{\sigma})(1-\frac{\rho( \mathbf{r})}{\rho_{\rm{sat}}})\delta( \mathbf{r}- \mathbf{r}'),
  \end{eqnarray}
for the particle-particle channel. The pairing strength $V_0=348.40$~MeV~fm$^{3}$ and energy cutoff $E_{\rm{cut}}^{\rm{q.p.}}=60$ MeV in the quasiparticle space are fitted for reproducing the proton pairing energy in the spherical nucleus $^{20}$Mg given by the RHB calculation with Gogny pairing force D1S~\cite{Berger1984NPA}.

Following the corresponding convergence check in Ref.~\cite{Li2012PhysRevC.85.024312}, the DDDRHB equations are solved in a spherical Dirac Woods-Saxon basis~\cite{Zhou2003PRC} determined by a box size $R_{\rm{max}}=20$ fm and a mesh size $\Delta r=0.1$ fm with the energy cutoff $E_{\rm{cut}}^{+}=150$ MeV and quantum number cutoff for angular moment $j_{\rm{max}}=19/2$.
For each $(l,j)$ block in Woods-Saxon basis, the number of negative energy states in the Dirac sea is the same as that of positive energy states.

Furthermore, the cutoffs are also necessary for the Legendre expansions of the densities, coupling strengths, meson fields, and potentials in the calculations. They are denoted with $\lambda_{\rho,\rm{max}}$, $\lambda_{g,\rm{max}}$, $\lambda_{\phi,\rm{max}}$, and $\lambda_{v,\rm{max}}$, respectively, where $\lambda_{\rho,\rm{max}}=\lambda_{v,\rm{max}}=4$ are used as in Refs.~\cite{Zhou2010Phys.Rev.C011301R,Li2012PhysRevC.85.024312}. In the following, we investigate the dependence of the DDDRHB results on $\lambda_{g,\rm{max}}$ and $\lambda_{\phi,\rm{max}}$ due to the density-dependent meson-nucleon couplings.

  \begin{figure}
      \includegraphics[width=8cm]{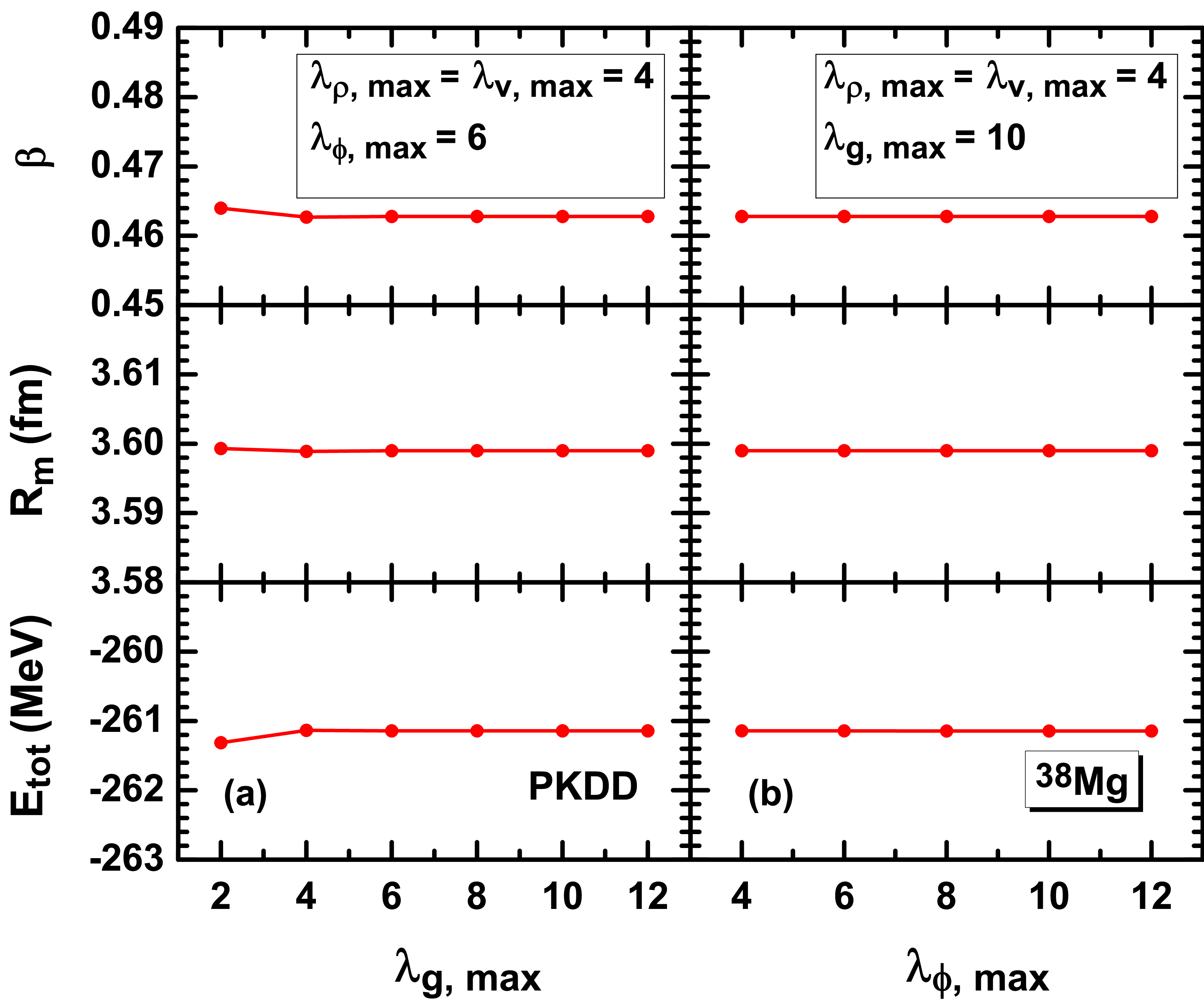}
      \caption{(Color online) Total energy $E_{\rm{tot}}$, matter rms radius $R_m$, and quadrupole deformation $\beta$ of $^{38}$Mg as functions of the cutoffs in Legendre expansions for (a) the coupling strengths $\lambda_{g, \rm{max}}$ and (b) the meson fields $\lambda_{\phi, \rm{max}}$, respectively.\label{fig1}}
  \end{figure}

In Fig.~\ref{fig1}, the total energy $E_{\rm{tot}}$, matter root mean square (rms) radius $R_m$, and quadrupole deformation $\beta$ are plotted as functions of the cutoffs in Legendre expansions for the coupling strengths $\lambda_{g, \rm{max}}$ and meson fields $\lambda_{\phi, \rm{max}}$, respectively. The convergence of these quantities with $\lambda_{g,\rm{max}}$ and $\lambda_{\phi,\rm{max}}$ is seen. The relative differences of $E_{\rm{tot}}$, $R_m$, and $\beta$ between the calculations with $\lambda_{g,\rm{max}}=10$ ($\lambda_{\phi,\rm{max}}=6$) and $\lambda_{g,\rm{max}}=12$ ($\lambda_{\phi,\rm{max}}=8$) are less than $0.001\%$. Therefore, the cutoffs $\lambda_{g,\rm{max}}=10$ and $\lambda_{\phi, \rm{max}}=6$ are adopted in the following calculations.

\begin{table}
\caption{Ground state properties of $^{38}$Mg calculated by RCHB~\cite{Meng1998Nucl.Phys.A3} as well as DDDRHB in the spherical case ($\lambda_{\rho, \rm{max}}=0$) and deformed case ($\lambda_{\rho, \rm{max}}=4$) with the functional PKDD~\cite{Long2004Phys.Rev.C034319}.
The rms radii for neutron $R_n$, proton $R_p$, and matter $R_m$ are in units of fm. The total energy $E_{\rm{tot}}$ and the corresponding contributions from the particles $E_{\rm{part}}$, meson fields $E_{\sigma}$, $E_{\omega}$, and $E_{\rho}$, Coulomb field $E_{\rm{coul}}$, rearrangement term $E_{\rm{rear}}$, pairing $E_{\rm{pair}}$, and center-of-mass correction $E_{\rm{cm}}$ are in units of MeV.
 }\label{table1}
\begin{ruledtabular}
\begin{tabular}{lrrr}
&\multicolumn{1}{c}{RCHB}
&\multicolumn{2}{c}{DDDRHB}\\
& & \multicolumn{1}{c}{$\lambda_{\rho, \rm{max}}=0$}
& \multicolumn{1}{c}{$\lambda_{\rho, \rm{max}}=4$}\\
\hline
$\beta$ &   &    & 0.46 \\
\hline
$R_n$   & 3.73 & 3.73  &3.80  \\
$R_p$   & 3.04 & 3.04  & 3.12 \\
$R_m$   & 3.53 & 3.53  &   3.60\\
\hline
$E_{\rm{part}}$& $-$856.36 & $-$856.32  &$-$877.05\\
$E_{\sigma}$   &  4886.42  & 4887.47    & 4919.31 \\
$E_{\omega}$   & $-$4070.70& $-$4071.61 &$-$4103.55\\
$E_{\rho}$     & $-$38.70  & $-$38.73   & $-$38.90\\
$E_{\rm{coul}}$& $-$32.20  & $-$32.20   & $-$32.02\\
$E_{\rm{rear}}$& $-$121.29 & $-$121.30  & $-$119.45\\
$E_{\rm{pair}}$& $-$14.53  & $-$14.57   & $-$1.67\\
$E_{\rm{cm}}$  & $-$7.25   & $-$7.25    & $-$7.80 \\
$E_{\rm{tot}}$ & $-$254.61 & $-$254.50  & $-$261.14\\
 \end{tabular}
\end{ruledtabular}
\end{table}

\begin{figure}
      \includegraphics[width=8cm]{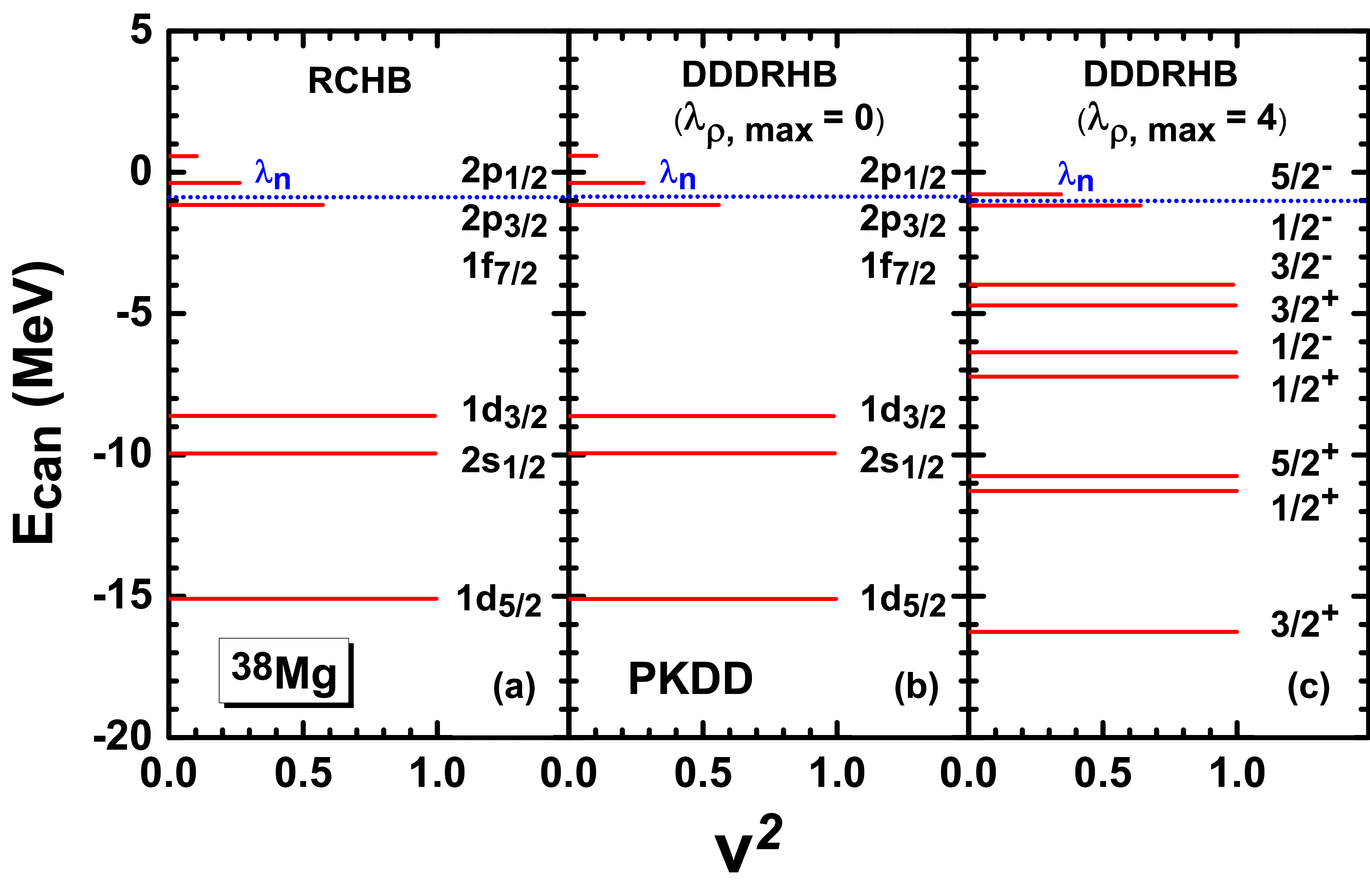}
      \caption{(Color online) Single-particle energies $E_{\rm{can}}$ of neutrons in the canonical basis above $-20$~MeV and their occupation probabilities $v^2$ of $^{38}$Mg calculated by (a) RCHB and by DDDRHB in the (b) spherical case ($\lambda_{\rho, \rm{max}}=0$) and (c) deformed case ($\lambda_{\rho, \rm{max}}=4$). The Fermi surfaces are shown as the dotted lines.}\label{fig2}
\end{figure}

In order to verify the accuracy of the present DDDRHB code, we first calculate the neutron-rich nucleus $^{38}$Mg with DDDRHB but constrained to the spherical case by taking $\lambda_{\rho, \rm{max}}=0$. The obtained bulk properties including the rms radii as well as the total energy and contributions from each component are listed in the middle column of Table~\ref{table1}. For comparison, the corresponding results calculated by the spherical relativistic continuum Hartree-Bogoliubov (RCHB)~\cite{Meng1998Nucl.Phys.A3} theory are shown in the left column. It is found that both calculations agree with each other quite well. The rms radii are the same up to 0.01~fm, and the total energies are the same up to 0.11~MeV which corresponds to an accuracy of $\sim0.04\%$.

In addition, the neutron single-particle energies in the canonical basis of $^{38}$Mg calculated by DDDRHB with $\lambda_{\rho, \rm{max}}=0$ are compared to those by RCHB in Fig.~\ref{fig2}. The length of each level is proportional to the occupation probability $v^2$. It can be seen that an excellent agreement is achieved.

In order to investigate the deformation effect, the neutron-rich nucleus $^{38}$Mg is calculated with DDDRHB by taking $\lambda_{\rho, \rm{max}}=4$ following the convergence study in Ref.~\cite{Zhou2006AIPConf.Proc.90}.
The corresponding bulk and neutron single-particle properties are also shown in Table \ref{table1} and Fig.~\ref{fig2}, respectively. In this case, the minimum in the potential energy surface locates at a prolate deformation with $\beta=0.46$, and the rms radii increase by $\sim2\%$. Due to the spherical symmetry breaking, the degenerate single-particle levels are split and labeled by the third component of the total angular moment and parity $\Omega^\pi$.

\begin{figure}
      \includegraphics[width=7cm]{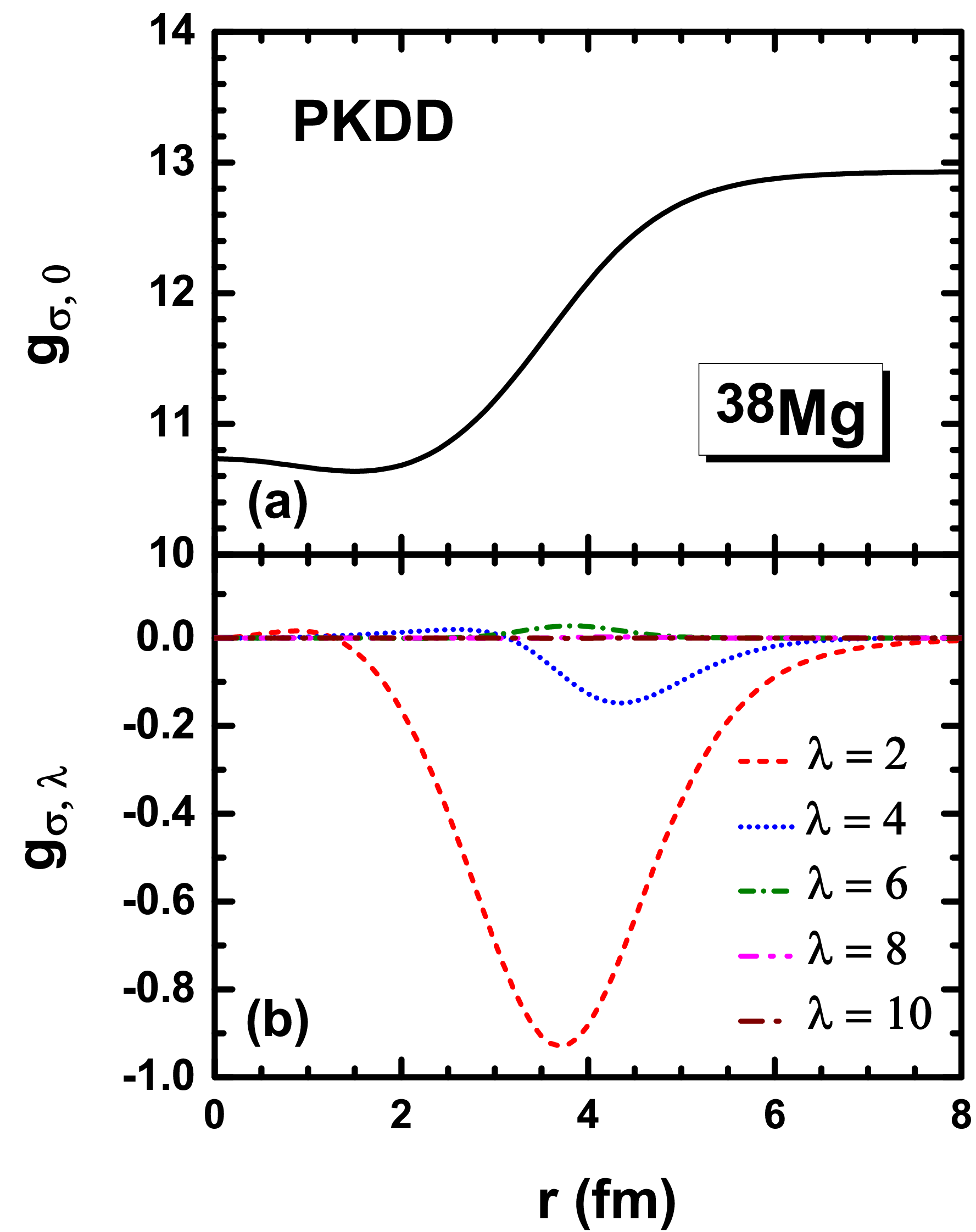}
      \caption{(Color online) Decompositions of coupling strength $g_{\sigma,\lambda}$ with (a) $\lambda=0$ and (b) $\lambda=2,4,6,8,10$ for $^{38}$Mg.}\label{fig3}
\end{figure}

In order to illustrate the partial waves of density-dependent coupling strengths, we show in Fig.~\ref{fig3} the $g_{\sigma,\lambda}$ in Eq.~(\ref{eq:gsig}) for $^{38}$Mg. It can be seen that the major component is that with $\lambda=0$, which is more than one order of magnitude larger than the others. The amplitudes of $g_{\sigma,\lambda}$ decrease quickly with increasing $\lambda$ and become negligible when $\lambda\geqslant8$. This confirms that the cutoff $\lambda_{g,\rm{max}}=10$ is reliable.

In summary, the density-dependent deformed relativistic Hartree-Bogoliubov theory in continuum is developed. The key formalism on handling the densities, coupling strengths, meson fields, and potentials in a deformed system with the partial wave method is presented. The newly developed DDDRHB code is verified by comparing the bulk and single-particle properties of the neutron-rich nucleus $^{38}$Mg calculated in the spherical case with $\lambda_{\rho,\rm{max}}=0$ to those obtained by the spherical RCHB. As an illustration, the nucleus $^{38}$Mg is also studied by DDDRHB in the deformed case with $\lambda_{\rho,\rm{max}}=4$. The corresponding minimum in the potential energy surface locates at a prolate deformation with $\beta=0.46$. In this case, the major components of density-dependent coupling strengths are those with $\lambda=0$, while the contributions from $\lambda=2,~4,~6$ components are not negligible.

\section*{ACKNOWLEDGMENTS}

This work was partially supported by the Major State 973 Program 2007CB815000, National Natural Science Foundation of China under Grants No. 10975008, No. 11105005, No. 11105006, No. 11175002, China Postdoctoral Science Foundation under Grants No. 20100480149, No. 201104031, and the Research Fund for the Doctoral Program of Higher Education under Grant No. 20110001110087.


%

\end{document}